\documentclass[sigconf,natbib=false]{acmart}

\AtBeginDocument{%
  }

\acmYear{2024}\copyrightyear{2024}
\setcopyright{acmlicensed}
\acmConference[HPDC '24]{The 33rd International Symposium on High-Performance Parallel and Distributed Computing}{June 3--7, 2024}{Pisa, Italy}
\acmBooktitle{The 33rd International Symposium on High-Performance Parallel and Distributed Computing (HPDC '24), June 3--7, 2024, Pisa, Italy}
\acmDOI{10.1145/3625549.3658896}
\acmISBN{979-8-4007-0413-0/24/06}

\RequirePackage[
  datamodel=acmdatamodel,
  style=acmnumeric,
  ]{biblatex}
\usepackage[frozencache,cachedir=.]{minted}
\usepackage{listings}
\hypersetup{
           breaklinks=true,   % splits links across lines
           colorlinks=true,   % displays links as colored text instead of blocks
           pdfusetitle=true,  % \title and \author values into pdf metadata
        }
\begin{document}

%%
%% The "title" command has an optional parameter,
%% allowing the author to define a "short title" to be used in page headers.
\title{Benchmarking Machine Learning Applications on Heterogeneous Architecture using Reframe}

%%
%% The "author" command and its associated commands are used to define
%% the authors and their affiliations.
%% Of note is the shared affiliation of the first two authors, and the
%% "authornote" and "authornotemark" commands
%% used to denote shared contribution to the research.
\author{Christopher Rae}
\orcid{0009-0001-2721-3562}
\email{crae@ed.ac.uk}
%\authornote{Both authors contributed equally to this research.}
\author{Joseph K. L. Lee}
\orcid{0000-0002-1648-2740}
\email{j.lee@epcc.ed.ac.uk}
%\authornotemark[1]
\author{James Richings}
\orcid{0000-0003-0895-4063}
\email{j.richings@epcc.ed.ac.uk}
\author{Mich\`{e}le Weiland}
\orcid{0000-0003-4713-3073}
\email{m.weiland@epcc.ed.ac.uk}
\affiliation{%
  \institution{EPCC, The University of Edinburgh}
  \streetaddress{EH8 9BT, UK}
  \country{UK}}

%%
%% By default, the full list of authors will be used in the page
%% headers. Often, this list is too long, and will overlap
%% other information printed in the page headers. This command allows
%% the author to define a more concise list
%% of authors' names for this purpose.
%\renewcommand{\shortauthors}{Rae, Lee, Richings}

%%
%% The abstract is a short summary of the work to be presented in the
%% article.
\begin{abstract}
With the rapid increase in machine learning workloads performed on HPC systems, it is beneficial to regularly perform machine learning specific benchmarks to monitor performance and identify issues. Furthermore, as part of the Edinburgh International Data Facility, EPCC currently hosts a wide range of machine learning accelerators including Nvidia GPUs, the Graphcore Bow Pod64 and Cerebras CS-2, which are managed via Kubernetes and Slurm. We extended the Reframe framework to support the Kubernetes scheduler backend, and utilise Reframe to perform machine learning benchmarks, and we discuss the preliminary results collected and challenges involved in integrating Reframe across multiple platforms and architectures.
\end{abstract}

%%
%% The code below is generated by the tool at http://dl.acm.org/ccs.cfm.
%% Please copy and paste the code instead of the example below.
%%


%%
%% Keywords. The author(s) should pick words that accurately describe
%% the work being presented. Separate the keywords with commas.

\received{27 March 2024}
%\received[revised]{12 March 2009}
%\received[accepted]{5 June 2009}

%%
%% This command processes the author and affiliation and title
%% information and builds the first part of the formatted document.
\maketitle

\section{Introduction}

To support the increasing trend of Machine Learning (ML) applications being used in scientific research and taking advantage of HPC systems, national HPC providers are adapting in two ways:
1) by providing compute services more suitable for data scientists which are more akin to cloud platforms, and 2) adoption of dedicated ML accelerators besides traditional HPC hardware (CPUs and GPUs). For example, at EPCC, besides managing HPC systems, ARCHER2 (the UK national HPC service) and Cirrus (an EPSRC tier-2 system), a new collection of data science-focused services that come under the umbrella of the Edinburgh International Data Facility (EIDF)~\cite{eidf} has been created to support research and data-driven innovation for academic and commercial partners. As part of the EIDF service, we offer access to a GPU cluster, large memory system (HPE Superdome Flex), as well as dedicated ML accelerators including the Cerebras CS-2~\cite{cerebras} and Graphcore Bow Pod64~\cite{graphcorePod64}. To use the EIDF GPU service and the Graphcore system, Kubernetes is used to manage workload, instead of a HPC scheduler such as Slurm~\cite{slurm}. This is to allow data scientists to run their applications in a reproducible and portable container environment, with maximum ease and minimal setup required. 

With the increasingly diverse range of services being offered, it is critical to regularly perform regression testing and benchmarking to maintain quality of service and detect issues. This allows us to monitor hardware performance and software compatibility, track performance around system changes or updates, to ensure service quality. The aim of this work is to create a framework for repeatable testing of multiple hardware architectures at EPCC. 

The main contributions of this paper are:
\begin{itemize}
    \item Integrate Kubernetes as a backend for the Reframe~\cite{reframe} testing framework, and explain the steps required to set up test cases
    \item Demonstrate and compare ML benchmarks (ResNet-50, DeepCam, CosmoFlow) on multiple EPCC systems including CPU, GPU, Cerebras CS-2, Graphcore Bow Pod64
    \item Discuss challenges in performing benchmarks on novel ML accelerators
\end{itemize}

For this work, we focus on benchmarks from the MLPerf Training~\cite{mlperf} and MLPerf HPC~\cite{mlperf_hpc} suite; however the framework can be easily ported to perform other benchmarks including popular LLM applications. This paper is structured as follows: in section~\ref{sec:background} we first introduce the technologies relevant for this work, then in section~\ref{sec:porting} we demonstrate how to set up a test case and some details behind the Kubernetes backend in Reframe. In section~\ref{sec:results} we show some preliminary results for benchmarks across multiple hardware architecture, and discuss some of the challenges involved in performing these benchmarks on novel systems.

\section{Background}
\label{sec:background}

\subsection{Reframe}
ReFrame~\cite{reframe} is a robust and portable framework developed by CSCS/ETH Zurich, designed for crafting system regression tests and benchmarks, with a particular focus on High Performance Computing (HPC) environments. Within ReFrame, tests are structured as Python classes, encapsulating the tests' variables and parameters. ReFrame orchestrates the loading and parallel execution of these tests through a structured pipeline. This pipeline manages every aspect of system interaction, encompassing tasks like switching programming environments, compiling code, submitting jobs, querying job statuses, as well as performing sanity checks and evaluating performance.

ReFrame supports a range of common job schedulers used by HPC systems such as Slurm, Flux~\cite{flux}, PBS~\cite{openpbs} and more. It is widely used by a number of International HPC centres, such as the ExCALIBUR Reframe~\cite{excalibur_reframe_paper,excalibur_reframe_repo} framework for multiple UK HPC sites, and within EPCC, it is run weekly to test the national supercomputer ARCHER2. One scheduler that is not supported by ReFrame is Kubernetes (K8s), which is starting to see usage within the intersection of data-scientist and HPC users, as is the case for EIDF's GPU service.

\subsection{Kubernetes}
Kubernetes (K8s)~\cite{k8s} is an open-source container orchestration system for automating software deployment and management of containerized applications. Originally designed by Google, the project is now maintained by the Cloud Native Computing Foundation. It is widely used by most cloud computing services, and a major benefit for users is that it abstracts away the underlying infrastructure such as networking, storage, and compute resources, and can therefore focus on developing the containerized applications which is guaranteed to be portable. For data scientists and ML practitioners, Kubernetes facilitates high scalability, which allows them to run more copies of a machine learning models at the same time, or add more compute resources (e.g. GPUs, RAM, cores) to train larger models. The containerized approach also ensures reproducibility and transparency, and is the common technology required for accessing cloud computing resources.

The basic scheduling unit within Kubernetes is a pod, which has a unique IP address and consists of one or more containers running on the same node. A container (which holds the running application, libraries, and dependencies) resides within a pod. There are additional higher level abstractions of workloads, such as Jobs, that can be defined and utilised, depending on the exact requirements. 

Kubernetes is not the only method to orchestrate and run containerized applications in data centers; in fact most HPC services already support running containers for example using Singularity~\cite{singularity}. However, Kubernetes can offer some advantages such as self-healing and automated fail-over, and dynamically adjusting resource utilisation to improve efficiency, which are beneficial to users who do not require the hardware characteristics of HPC systems including high-speed interconnect and I/O systems. For the EIDF service at EPCC, Kubernetes is used to access and manage the GPU service and the Graphcore system. The integration of Kubernetes as a backend for Reframe in this work allows us to use the same framework to test multiple systems managed by EPCC.

\subsection{ML Benchmarks}
To demonstrate the use of Reframe with Kubernetes on our services, and for future monitoring of performances, we utilise three main benchmarks: ResNet-50~\cite{resnet}, DeepCam~\cite{deepcam}, and CosmoFlow~\cite{cosmoflow}. ResNet-50 is a residual network designed for image classification, and the ResNet-50 v1.5 is included as part of the MLPerf Training~\cite{mlperf} suite. CosmoFlow is a 3D convolutional neural network trained on N-body cosmology simulation data, and is used for predicting cosmological parameters from the distribution of dark matter in the universe; DeepCAM implements a convolutional encoder-decoder segmentation architecture (deeplabv3plus\_xception~\cite{deeplabv3plus}) trained on CAM5 climate simulation data with TECA~\cite{TECA} generated heuristic segmentation masks to identify extreme weather phenomena, such as atmospheric rivers and tropical cyclones. DeepCAM was the first deep learning application which scaled to the full OLCF Summit system, and with CosmoFlow, are part of the MLPerf HPC~\cite{mlperf_hpc} suite.

The MLPerf Training and MLPerf HPC suites are managed by MLCommons, which are used to monitor and compare machine learning performance across different architectures. We have implemented the models using Pytorch~\cite{pytorch} which can be found at \url{https://github.com/EPCCed/reframe-mlperf-epcc}. We have also attempted to port the benchmarks to the Graphcore and Cerebras system, which requires using the provided compiler and libraries with varying degree of support for the Pytorch functionalities. We managed to run ResNet-50 on both systems, and CosmoFlow with half-precision on Graphcore; the difficulties we encountered will be described in more detail in section~\ref{sec:challenges-discussions}.

\subsection{Hardware Architecture}
\begin{table*}[!h]
    \caption{Hardware targets of this study, the scheduling solution, communication backend, and associated file system}
    \label{tab:hardware-architecture}
    %\centering
    \begin{tabular}{l|l|l|l|l|l}
    \toprule
        Service & Hardware & Scheduler & Launcher & Communication Backend & File System Type \\ 
    \midrule
        ARCHER2~\cite{archer2_hardware} &  AMD EPYC 7742 64-core & Slurm & srun & MPI~\cite{mpi} & Lustre \\ 
        ARCHER2 & AMD MI210 & Slurm & torchrun & RCCL~\cite{rccl} & Lustre \\ 
        Cirrus~\cite{cirrus_hardware} & Nvidia V100 & Slurm & srun & MPI & Lustre \\ 
        EIDF~\cite{eidf_gpuservice} & Nvidia A100 & Kubernetes & torchrun & NCCL~\cite{nccl} & Ceph \\ 
        EIDF & Graphcore Bow Pod64 & Kubernetes & PopRun & IPU-Fabric & Ceph \\ 
        EIDF & Cerebras CS-2 & Slurm & None & SwarmX & Lustre \\ 
    \bottomrule
    \end{tabular}
\end{table*}

Table \ref{tab:hardware-architecture} lists some of the services managed by EPCC, and the specific hardware model. As previously mentioned, the EIDF service~\cite{eidf_gpuservice} includes access to Nvidia GPUs (H100 and A100), Graphcore Bow Pod64, and Cerebras CS-2; the first two utilise Kubernetes to orchestrate workloads, and the Cerebras CS-2 is managed using Slurm. The Graphcore and Cerebras systems are both highly parallel accelerators with high bandwidth, designed specifically for AI/ML workoads; a Bow Pod64~\cite{graphcorePod64} system contains 94,208 individual IPU Cores across 64 Bow IPUs, and a Cerebras CS-2~\cite{cerebras} contains 850,000 cores on a single wafer. We also compare it against two HPC systems, ARCHER2 and Cirrus, both of which use Slurm for scheduler. ARCHER2~\cite{archer2_hardware} is an HPE Cray EX system, with two AMD EPYC Rome 7742 CPUs per node, and a smaller GPU development platform with 4 AMD MI210 GPUs per node. Cirrus~\cite{cirrus_hardware} consists of a GPU partition, with 4 Nvidia V100 GPUs per node.

For this work, we utilise Reframe with the Kubernetes backend, which will be introduced in section~\ref{sec:porting}, to compare the performance of the ML benchmarks across multiple hardware platforms, in particular, comparing novel hardware against traditional HPC architecture.

\section{Porting Reframe for Kubernetes}
\label{sec:porting}
In this section we demonstrate an example of how to set up a test case, and describe how the Kubernetes backend was integrated into the Reframe framework, as well as some extensions that can be added. The implementation of this work is open-source and available at \url{https://github.com/BigBalloon8/reframe}.

\subsection{Example}
To write your K8s test you will first need to define your configuration:
\begin{minted}{python}
site_configuration = {
  "systems": [
    {
      "name": "eidf",
      ...
        "partitions": [
          {
            "name": "gpu-service",
            "scheduler": "k8s",
            "launcher": "k8s",
            ...
          },
       ],
    }
  ],
  "environments": ...,
  "logging": ...
}
\end{minted}
The scheduler and launcher are set to K8s to enable the Kubernetes scheduler backend.

Next, to define the workload, the user will have to set up a \texttt{yaml} file to configure the container and application. The example shown here is a simplified version of the test for running the ResNet-50 benchmark on the A100 GPUs on the EIDF GPU Service:

\begin{figure}[H]
\begin{minted}{yaml}
#/path/to/resnet50_pod.yml
apiVersion: v1
kind: Pod
metadata:
  name: 'ResNet50-Test'
spec:
  restartPolicy: Never
  containers:
    - name: 'resnet-test'
      image: bigballoon8/mlperf-epcc 
      workingDir: '/workspace/ML/ResNet50/Torch'
      command:
        - torchrun
      args:
        - "--nproc_per_node=4"
        - "train.py"
        - "-c /workspace/ML/ResNet50/Torch/config.yaml"
      resources:
        limits:
          cpu: 16
          memory: 32Gi
          nvidia.com/gpu: '4'
      volumeMounts:
        - mountPath: /mnt/ceph_rbd
          name: volume
  nodeSelector:
    nvidia.com/gpu.product: 'NVIDIA-A100-SXM4-40GB'
  volumes:
    - name: volume
      persistentVolumeClaim:
        claimName: 'imagenet-pvc'
\end{minted}
\end{figure}
To set up and launch the Kubernetes workload from Reframe, there are two ways to do so: 1) pass the path to the config yaml as a path-like string, or 2) the config can be read inside of the test and passed as a Python container with the contents of the config within it:

\begin{minted}{python}
@rfm.simple_test
class ResNet50Test(rfm.RunOnlyRegressionTest):
  valid_systems = ['eidf:gpu-service']
  valid_prog_environs = ["*"]
  k8s_config = "/path/to/resnet50_pod.yml"

# OR

@rfm.simple_test
class ResNet50Test(rfm.RunOnlyRegressionTest):
  valid_systems = ['eidf:gpu-service']
  valid_prog_environs = ["*"]

  @run_after("init")
  def k8s_setup(self):
    k8s_config_path = "/path/to/resnet50_pod.yml"
    with open(k8s_config_path, "r") as stream:
        pod_info = yaml.safe_load(stream)
     self.k8s_config = pod_info
     
\end{minted}

While both options lead to the same result, the second option can be used to create parameterised tests:

\begin{minted}{python}

# k8s_pod_test.py
@rfm.simple_test
class ResNet50Test(rfm.RunOnlyRegressionTest):
  valid_systems = ['eidf:gpu-service']
  valid_prog_environs = ["*"]
  num_gpus = parameter([4, 8])

  @run_after("init")
  def k8s_setup(self):
    k8s_config_path = "/path/to/cuda-pod.yml"
    with open(k8s_config_path, "r") as stream:
      pod_info = yaml.safe_load(stream)
    
    pod_info["spec"]["containers"][0]["args"] = [
      f"--nproc_per_node={self.num_gpus}", 
      "train.py", 
      "-c /workspace/ML/ResNet50/Torch/config.yaml",
    ]
    pod_info["spec"]["containers"][0]\
    ["resources"]["limits"]\
    ["nvidia.com/gpu"] = self.num_gpus
    self.k8s_config = pod_info
\end{minted}

\subsection{Behind The Scenes}
We introduced an extension to ReFrame that allows users to write regression tests and benchmarks for K8s clusters. This is done by implementing a custom scheduler that can interact with the K8s API. The scheduler role can be split into 4 steps:

\begin{enumerate}
    \item Launch the K8s workload
    \item Wait for all pods associated with the workload to finish
    \item Write the logs of all the pods associated with the workload to an output file (\texttt{rfm\_job.out}) 
    \item Clean Up the workload
\end{enumerate}

For the scheduler to identify the individual workload resources, a unique 8-character long random string is generated for each reframe test. The scheduler reads the k8s\_config and updates all of the metadata attributes found in the config by appending \\ \texttt{\{"rfm":identifier\}} to the metadata's labels. The identifier is used to identify both which workload resources are associated with ReFrame and which are associated with the specific test, and this allows for multiple tests to run in parallel. 

Next, the K8s workload is launched (similar to manually launching via \texttt{kubectl create -f /path/to/k8s\_config.yaml}). Upon launching, a logging thread will be generated. The job of this thread is to write the output of all the pods associated with the workload to the \texttt{stdout} of the test until all the pods have either succeeded, failed, or crashed. The thread can identify which pods are associated with the given test by the unique \texttt{identifier}.

Once the workload has been launched the scheduler will wait for one of three events:
\begin{enumerate}
    \item All pods associated with the workload of the test to succeed, fail or crash
    \item The test's time limit is reached
    \item The user cancels the test
\end{enumerate}
Once one of the above events has happened the scheduler will check to see if all the associated pods completed successfully, if so the scheduler will wait for the logging thread to complete and clean up the workload. If one or more pods are unsuccessful, the scheduler will terminate the logging thread and print the logs, while keeping the workload resources active (i.e. not cleaning up) to allow the users to manually inspect the pods or workload and assess the problem. If the user cancels the test, the scheduler will close the logging thread, print the logs to the \texttt{stdout} and automatically clean up the workload resources.

Other than Pods, other workload resources are also supported. For Jobs, the scheduler will dynamically extract the workload resource's type; if the scheduler detects that the workload resource is a Job, the scheduler will read the completions value found in the job spec of the k8s\_config or set it to 1 if it is not provided. The scheduler will then wait for the predefined number of completed pods associated with that test Job to finish. Besides Jobs, other workload resources, e.g. IPUJobs for Graphcore workloads, are also available as experimental features, which is described in the documentation.

ReFrame also allows you to specify certain global command-line options used by K8s, including the namespace and the context. By default the namespace will be set to the \texttt{default} namespace and the context will use the current context defined in the environment variable \texttt{KUBECONFIG}. These can be set as options within the regression test:

\begin{minted}{python}
@rfm.simple_test
class ResNet50Test(rfm.RunOnlyRegressionTest):
    valid_systems = ['eidf:gpu-service']
    valid_prog_environs = ["*"]
    k8s_config = "/path/to/resnet50_pod.yml"

    namespace = "NAMESPACE"
    context = "CONTEXT"
\end{minted}

\section{Benchmarking Results}
\label{sec:results}

\begin{table*}[!ht]
    \centering
    \captionsetup{justification=centering}
    %TODO update caption
    \caption{Througuhput for training ResNet-50 on ImageNet1k, using a global batchsize of 32}
    \label{tab:ResNet50 Performance}
    \begin{tabular}{c|c|c|c|c}
    \toprule
        Hardware & No. Processing Units & Compute Throughput (inputs/s) & Compute Fraction & Effective Throughput (input/s)\\ 
    \midrule
        ARCHER2 CPU & 4 CPU& 40.5 & 98.9\% & 40.1 \\ 
        ARCHER2 MI210 & 4 GPU& 293.2 & 77.3\% & 226.6 \\ 
        Cirrus V100 & 4 GPU& 138.0 & 97.3\% & 134.3 \\ 
        EIDF A100 & 4 GPU& 226.2 & 79.4\% & 179.7 \\ 
        Graphcore & 8 IPU& - & - & 255.6\\ 
        Cerebras CS-2 & 1 WSE& - & - & 452.0\\ 
    \bottomrule
    \end{tabular}
    \centering
\end{table*}

\begin{table*}[!ht]
    \centering
    \captionsetup{justification=centering}
    %TODO update caption
    \caption{Throughput for the CosmoFlow benchmark, using a global batchsize of 32}
    \label{tab:CosmoFlow Performance}
    \begin{tabular}{c|c|c|c|c}
    \toprule
        Hardware & No. Processing Units & Compute Throughput (inputs/s)& Compute Fraction & Effective Throughput (input/s) \\ 
    \midrule
        ARCHER2 CPU & 4 CPU&  14.9 & 98.9\% & 14.8\\ 
        ARCHER2 MI210 & 4 GPU&  479.9 & 15.1\% & 72.5\\ 
        Cirrus V100 & 4 GPU&  112.2 & 69.6\% & 78.1\\ 
        EIDF A100 & 4 GPU&  117.9 & 49.3\% & 58.1\\ 
        Graphcore (half precision) & 8 IPU&  -  & - & 14.5 \\ 
    \bottomrule
    \end{tabular}
    \centering
\end{table*}

\begin{table*}[!ht]
    \centering
    \captionsetup{justification=centering}
    %TODO update caption
    \caption{Throughput for the DeepCAM benchamrk, using a global batchsize of 32}
    \label{tab:DeepCAM Performance}
    \begin{tabular}{c|c|c|c|c}
    \toprule
        Hardware & No. Processing Units & Compute Throughput (inputs/s) & Compute Fraction & Effective Throughput (input/s)\\ 
    \midrule
        ARCHER2 CPU & 8 CPU &  6.1  & 98.7\% & 6.1\\ 
        ARCHER2 MI210 & 4 GPU&  26.4  & 55.0\% & 14.5\\
        Cirrus V100 & 4 GPU& 54.1  & 54.1\% & 15.4 \\ 
        EIDF A100 & 4 GPU& 101.7 & 13.5\% & 13.7 \\
    \bottomrule
    \end{tabular}
    \centering
\end{table*}

For each benchmark, we use Reframe to capture the time spent on computation and I/O per epoch, and the throughput (number of inputs processed per second). These metrics are more important for comparing across hardware types, and also to track variation of hardware performance over time. The time to quality metric typically used for MLPerf benchmarks requires running to completion, which takes a significantly long time, and is highly stochastic and variable between runs, is therefore less useful for our purpose. 

Tables~\ref{tab:ResNet50 Performance}, \ref{tab:CosmoFlow Performance}, \ref{tab:DeepCAM Performance} summarise the performance for ResNet-50, CosmoFlow, and DeepCAM respectively. For each epoch, we measure the I/O time (duration spent copying data to GPU), and the remaining compute time. The \textit{Compute Throughput} is defined as the number of input items processed divided by the time spent on computation. This is useful for understanding the raw computational performance between the hardware architecture without the I/O performance dominating, which is often significant depending on the dataset and the underlying parallel file system. This can be measured from the \textit{Compute Fraction}, which is the time spent on computation relative to the total time per epoch. The \textit{Effective Throughput} is the number of items processed divided by the total time, which provides a complete picture of the performance of the whole system. 

On ARCHER2 CPU nodes, we ran ResNet-50 and CosmoFlow on 2 nodes (4 CPU sockets) with 16 MPI ranks per node, and DeepCAM on 4 nodes (8 CPU sockets) with 8 MPI ranks per node due to the larger memory requirement. For GPU runs (AMD MI210 on ARCHER2, Nvidia V100 on Cirrus, and Nvidia A100 on EIDF GPU Service), all tests were run on 4 GPUs. We were able to compile and run ResNet-50 and CosmoFlow on Graphcore, and these were executed on 8 IPUs. We also ran ResNet-50 on the a single Cerebras CS-2 machine. For each benchmark, a global batch size of 32 was used; for this initial study we are not focusing on tuning optimal parameters for the different hardware and instead have chosen the same set of parameters for all hardware. In fact, Reframe could be used here for parameterizing tests and performing hyperparameter tuning.

\subsection{Discussion}
From these preliminary results, we observed that the compute fraction on the different GPU systems vary significantly depending on the model/dataset, as well as the filesystem. In general, the compute fraction for ResNet-50 is much higher than CosmoFlow and DeepCAM, due to the fact that the deeper network requiring more compute per sample. We also observed that the performance of the A100 on EIDF to be slower than the older V100 GPUs on Cirrus; this could be attributed to the slower filesystem available to EIDF, and the fact that each node is shared with other users. This shows that the filesystem and I/O performance is hugely influential to the training speed, which may be more significant than the gains between GPU generations.

The low GPU utilisation and I/O bottleneck observed in DeepCAM and CosmoFlow presents a significant opportunity for optimisation. There are data-loading strategies which can improve this, for example by preloading the dataset onto the on-node storage or memory, which is done for most MLPerf HPC timing runs~\cite{mlperf_hpc}.

For the CPU runs, it is crucial to increase the MPI ranks in order to maximize the CPU utilisation. Nevertheless, the effective throughput per CPU is still significantly slower than that per GPU. However, we note that it is difficult to make a fair comparison of performance across platforms, due to difference in other components including filesystem and networking. 

\subsection{Challenges}
\label{sec:challenges-discussions}
For each MLPerf benchmark, a reference implementation is provided; the reference implementations for ResNet-50 and CosmoFlow use TensorFlow2~\cite{tensorflow}. However, only Pytorch is supported on the Cerebras system, which led us to move away from MLCommon's reference implementation and rewrite the models in Pytorch for a fairer comparison across hardware. Unfortunately, we encountered a number of issues when porting and running these benchmark models on the Graphcore and Cerebras machines.

The three benchmarks chosen are convolutional neural network (CNN) models, which are important for computer vision as well as analysing a wide range of 3D simulations. ResNet-50 is a deeper network processing small 2D data samples (average size of 124KB per input), whereas CosmoFlow and deepCAM are relatively shallow 3D CNN models, with much larger data samples (2.8MB and 61MB per input respectively).

The main challenge for porting to Graphcore is to work around the limited memory on each IPU (0.9GB). This require extensive use of pipeline parallelism, which is easy to implement on Graphcore with minimal changes to the original model definition. For our implementation of ResNet-50, we split half the convolution blocks on the first IPU and the other half on another IPU. For CosmoFlow, we were unable to split the pipeline stages onto the memory, which could be due to the large memory required for the activations of the first convolution block not fitting on a single IPU. A work-around was to train using half-precision, which allowed us to fit the model onto the system. We were unable to fit and compile the DeepCAM model without significant modification. 

In general, obtaining high performance, or even compiling complex models at all, on Graphcore requires optimising pipeline placement. This process is commonplace and simpler for Large Language Models (LLM), which tend to have symmetric blocks with the same amount of computational cost that can be pipelined evenly; this tends not to be true for CNNs, which will require careful partitioning of blocks onto separate IPUs. Performing these optimisations is beyond the goal of preliminary benchmarking and demonstration of the testing framework of this paper, but something we are interested in pursuing further. In fact, an implementation of ResNet-50 optimised for Graphcore IPUs with higher throughput is available \cite{graphcore_resnet}.

On the other hand, the Cerebras software stack is relatively rigid, and developers have less control over customisations of the training process and model development. We were able to port and run ResNet-50 on Cerebras without much issues, and observed very strong performance out of the box. However, for CosmoFlow, the issue we encountered is that as of writing, one of the fundamental operations in the model, \texttt{MaxPool3d}, is not supported. We attempted to emulate the same operation using a series of \texttt{MaxPool2d} operations, but exceeded the memory limit and failed to compile. Similarly, we were unable to compile the model for DeepCAM \\(\texttt{deeplabv3plus\_xception}) on Cerebras, where we ran into compilation errors. 

In general, the support for LLM models is better relative to CNN models on the Cerebras system (at the time of writing, the majority of model examples provided on the official modelzoo~\cite{cerebras_modelzoo} are LLMs, with 1 diffusion model DiT~\cite{DiT} and 1 multilayer perceptron model). We would like to see more support for CNNs, which will be a crucial component for development of multi-modal models in the future.

\section{Conclusion}
We have extended the Reframe testing framework to support Kubernetes as a backend scheduler, and we have utilised this Reframe backend to perform ML benchmarks (including ResNet-50, DeepCAM, and CosmoFlow from the MLPerf training and HPC suites) on a range of services and systems at EPCC, including different generations of Nvidia GPUs and the Graphcore Bow Pod64. 

This is an important capability for our centre, as it allows us to perform repeatable performance testing of ML workloads and to monitor application performance stability, which can assist in identifying performance impact due to system changes or upgrades etc. This will also allow us to directly compare performance across multiple services and architectures, and parametrise testing with different systems with a similar interface in an automated manner. 

We believe this tool will be valuable for other HPC centres, where with minimal update this could be used to perform regular ML benchmarking and performance testing straight away. It is also easy to extend the framework to support other Kubernetes workloads specific to a centre, and to add other ML applications and benchmarks. For example, other than benchmarking and monitoring ML application performance, we also use the Reframe framework to monitor the data transfer bandwidth of the Kubernetes Persistent Volumes, and perform other HPC tests including n-body simulation benchmarks on GPUs. This work is publicly available at \url{https://github.com/BigBalloon8/epcc-reframe} and a pull request will be submitted to update the EPCC Reframe repository.

%%
%% The acknowledgments section is defined using the "acks" environment
%% (and NOT an unnumbered section). This ensures the proper
%% identification of the section in the article metadata, and the
%% consistent spelling of the heading.
\begin{acks}
This work used the ARCHER2 UK National Supercomputing Service (https://www.ARCHER2.ac.uk). This work used the Cirrus UK National Tier-2 HPC Service at EPCC (http://www.cirrus.ac.uk) funded by the University of Edinburgh and EPSRC (EP/P020267/1). This work was supported by the Edinburgh International Data Facility (EIDF) and the Data-Driven Innovation Programme at the University of Edinburgh. The work presented in this paper was partially funded through the UKRI AI for Net Zero grant EP/Y004450/1. For the purpose of open access, the author has applied a Creative Commons Attribution (CC BY) licence to any Author Accepted Manuscript version arising from this submission.
\end{acks}

%%
%% Print the bibliography
%%
\printbibliography

%%
%% If your work has an appendix, this is the place to put it.
\appendix

\end{document}